\documentclass[10pt,draft]{dis03}
\usepackage{epsf,amsmath}

\def\z#1{{\zeta_{#1}}}

\def\nf{{n^{}_{\! f}}}
\newcommand{\MSb}{$\overline{\mbox{MS}}$}
\newcommand{\as}{\alpha_{\rm s}}

\newcommand{\bea}{\begin{eqnarray}}
\newcommand{\eea}{\end{eqnarray}}

\begin{document}
\noindent

\title{THREE-LOOP RESULTS AND SOFT-GLUON EFFECTS IN DIS}

\author{S. Moch$^{\, a}$, J.A.M. Vermaseren$^{\, b}$ and A. Vogt$^{\, b}$\\[2ex]
{\it $^a$Deutsches Elektronensynchrotron DESY }\\
{\it Platanenallee 6, D--15738 Zeuthen, Germany}\\[2ex]
{\it $^b$NIKHEF Theory Group} \\
{\it Kruislaan 409, 1098 SJ Amsterdam, The Netherlands}\\[2ex]}

\maketitle

\begin{abstract}
\noindent 
We have calculated the fermionic contributions to the flavour non-singlet 
structure functions in deep-inelastic scattering (DIS) at third order of 
massless perturbative QCD. We discuss their implications for the 
threshold resummation at the next-to-next-to-leading logarithmic accuracy.
\end{abstract}

\section{Introduction}
The calculation of perturbative QCD corrections for deep-inelastic 
structure functions is an important task.
The present and expected future experimental precision, 
for instance at HERA, calls for complete next-to-next-to-leading order
(NNLO) predictions.
These offer the possibility to determine the strong coupling constant 
$\alpha_s$ and to analyze the proton structure and its parton
content with unprecedented precision.
Knowledge of the latter is of particular importance for 
the analysis of hard scattering reactions at 
future LHC experiments.

At present, this level of accuracy is not yet possible, 
because the necessary anomalous dimensions governing 
the parton evolution at NNLO are not fully known. 
The two-loop coefficient functions of $F_2,F_3$ and
$F_L$ have been calculated some time 
ago~\cite{vanNeerven:1991nn,Zijlstra:1991qc,Zijlstra:1992kj,Zijlstra:1992qd,%
Moch:1999eb}, 
but for the corresponding three-loop anomalous dimensions 
$\gamma_{\rm pp'}^{(2)}$, only a finite number of fixed Mellin
moments~\cite{Larin:1994vu,Larin:1997wd,Retey:2000nq} 
are available.
As a first step towards the complete calculation, we have computed the 
fermionic three-loop contributions to the flavour non-singlet (NS) structure 
functions $F_2$ and $F_L$ in unpolarized electromagnetic deep-inelastic
scattering~\cite{Moch:2002sn}. 
Already these results have immediate consequences for threshold resummation of
soft gluons which we will discuss in the following. 

\section{Threshold resummation}

It is well known that perturbative QCD corrections to structure
functions receive large logarithmic corrections, which originate from 
the emission of soft gluons. These corrections are relevant at large
values of the scaling variable $x$ (in Mellin space at large values of
the Mellin moment $N$) and can be resummed to all orders in
perturbation theory. It is interesting to investigate the implications 
of our three-loop results~\cite{Moch:2002sn} for the threshold 
exponentiation~\cite{Sterman:1987aj,Catani:1989ne,Catani:1991rp} 
at next-to-next-leading logarithmic (NNLL) accuracy~\cite{Vogt:2000ci}.

Here the quark coefficient function for $F^{}_{2}$ can, up to terms 
which vanish for $N \to \infty$, be written as
\bea
\label{eq:csoft}
  C_{\,2}(\as, N) \: =\: 
  (1 + a_{\rm s\,} g_{01}^{} + a_{\rm s\,}^2 g_{02}^{} + \ldots ) \: 
  \exp \, [\,G^N(Q^2)  \, ] \, ,
\eea
where the resummation exponent $G^N$ can be expanded as 
\begin{equation}
\label{eq:GNexp}
  G^N(Q^2) \: =\:  
  L\, g_1^{}(a_{\rm s\,}L) + g_2^{}(a_{\rm s\,}L) + 
  a_{\rm s}\, g_3^{}(a_{\rm s\,}L) + \ldots\, 
\end{equation}
with $a_{\rm s} = \as/(4\pi)$ and $L = \ln N$. The functions 
$g_l^{}$ depend on universal coefficients $A_{\,i\,\leq\, l}$ and 
$B_{\,i\,\leq\, l\!-\!1}$ 
and process-dependent parameters $D_{\,i\,\leq\, l\!-\!1}^{\,\rm
DIS}$ (see e.g. ref.~\cite{Vogt:2000ci} for the precise definitions of the 
functions $g_{1,2,3}^{}$).

In a physical picture, the resummation of the perturbative expansion
for $C_{\,2}$ rests upon the refactorization of 
$C_{\,2}$ (valid in the threshold region of phase
space) into separate functions of the jet-like, soft, and off-shell
quanta that contribute to its quantum corrections. 
Each of the functions organizes large corrections
corresponding to a particular region of phase space~\cite{Contopanagos:1997nh}.

To NNLL accuracy, the function $g_3$ involves the new coefficients $A_3$,
$B_2$ and $D_2^{\,\rm DIS}$. These coefficients can be fixed by
expanding eq.~(\ref{eq:csoft}) in powers of $\as$ and comparing to the 
result of the full fixed-order calculation~\cite{Moch:2002sn}. 
In the \MSb\ scheme, the parameter $A_3$ is simply the coefficient 
of $\ln N$ in $\gamma_{\,\rm ns}^{(2)}(N)$ or, equivalently, 
of $1/(1-x)_+$ in $P_{\,\rm ns}^{(2)}(x)$. It reads
\bea
\lefteqn{
   A_3\: = \: (1178.8 \pm 11.5) }
\\ 
&&\nonumber
   + C_A C_F \nf\, \left[ - \frac{836}{27} + \frac{160}{9}\:\z2 
               - \frac{112}{3}\:\z3 \right]
   + C_F^{\,2} \nf\, \left[ -  \frac{110}{3}  + 32\:\z3 \right]
\\ 
&&\nonumber
   + C_F \nf^{\!\!2} \left[ - \frac{16}{27}\,\right] \:\: ,
\eea
where $C_A=3$, $C_F=4/3$ and $\nf$ is the number of light (massless)
flavours. The estimate of the non-fermionic part~\cite{Vogt:2000ci} is based on 
the approximations of $P_{\,\rm ns}^{(2)}(x)$ constructed in 
ref.~\cite{vanNeerven:2000wp} using the first six even-integer 
moments~\cite{Retey:2000nq} and its small-$x$ limit~\cite{Blumlein:1995jp}. 
The exact fermionic part has been obtained independently in 
refs.~\cite{Moch:2002sn,Berger:2002sv}, while the $\nf^{\! 2}$ contribution is 
already known from ref.~\cite{Gracey:nn}.

The complete results for $B_2$ and $D_2^{\,\rm DIS}$ can be 
inferred from fermionic result of the three-loop coefficient function 
$c_{\,2,\rm ns}^{\,(3)}$ in ref.~\cite{Moch:2002sn}, yielding
\bea
\label{eq:B2D2}
  B_2 \: & =\! &
  C_F^{\,2}  \left[ -\frac{3}{2} - 24\,\zeta_3 + 12\,\zeta_2 \right] 
  \: +\: C_F C_A  \left[ - \frac{3155}{54} + 40\,\zeta_3
                 + \frac{44}{3}\,\zeta_2 \right] 
  \\ & & \mbox{} + \:
  C_F \nf\,  \left[ \frac{247}{27} - \frac{8}{3}\,\zeta_2 \right] 
  \:\: ,   \nonumber\\
  D_2^{\,\rm DIS} \!\! & =\! & 0 \:\: .
\eea
As a matter of fact, the contribution to $c_{\,2,\rm ns}^{\,(3)}$ involves
only a linear combination, $\beta_0 (B_2 + 2\,D_2^{\,\rm DIS})$, with 
$\beta_0$ being the one-loop coefficient of the QCD $\beta$-function.
However, the different combination $B_2 + D_2^{\,\rm DIS}$ has been determined 
in ref.\cite{Vogt:2000ci} by comparing the expansion of eq.~(\ref{eq:csoft})
to the two-loop coefficient function $c_{\,2,\rm ns}^{\,(2)}$ of 
ref.~\cite{Matsuura:1989sm}.
Thus, $B_2$ and $D_2^{\,\rm DIS}$ can be disentangled. 
It is interesting to observe the vanishing of $D_1^{\,\rm DIS}$ and
$D_2^{\,\rm DIS}$, for which an all-order generalization 
has been proposed in ref.~\cite{Forte:2002ni,Gardi:2002xm}.
This is in contrast to the Drell-Yan process, 
where the functions $D_l^{\,\rm DY}$ are generally different 
from zero. 
For instance, $D_2^{\,\rm DY}$ has been derived in ref.~\cite{Vogt:2000ci}. 

Let us briefly illustrate numerically for large $N$ the improvement due to the NNLL 
corrections for the soft gluon exponent $G^N$ of deep-inelastic scattering.
In fig.~1 on the left, we show the resummation exponent $G^N(Q^2)$ of
eq.~(\ref{eq:GNexp}). Here, we choose $\mu_r^2 = \mu_f^2 = 
Q^2$, $\nf = 4$ and $\as = 0.2$, which corresponds to scales between 
about 25 and 50 GeV$^2$, depending on the precise value of $\as(M_Z^2)$. 
In fig.~1 on the right, we display the convolution with a schematic, 
but typical input evaluated with the so-called `minimal-prescription' 
contour~\cite{Catani:1996yz}. 
It is obvious from both figures that knowledge of the leading
logarithmic (LL) and next-to-leading logarithmic (NLL) 
terms~\cite{Sterman:1987aj,Catani:1989ne,Catani:1991rp} alone,
i.e., those enhanced by factors $\ln N (\alpha_s 
\ln N)^n$ and $(\alpha_s \ln N)^n$, 
is not sufficient for reliably determining the function $G^N$
and its impact after convolution even for rather large values of $N$ and
$x$. 

The NNLL corrections discussed here are rather small over a wide
range. This indicates that the soft-gluon exponent 
$G^N(Q^2)$ stabilizes and that the soft-gluon effects on the 
\MSb\ quark coefficient function can be reliably estimated.
Recall, however, that the NNLL corrections are large for the `physical
kernel' governing the scaling violations of the non-singlet structure
function $F_{2,\rm ns}$ \cite{vanNeerven:2001pe}.

\begin{figure}[!thb]
\vspace*{0.4cm}
\begin{center}
\centerline{
\epsfxsize=6.2cm\epsfbox{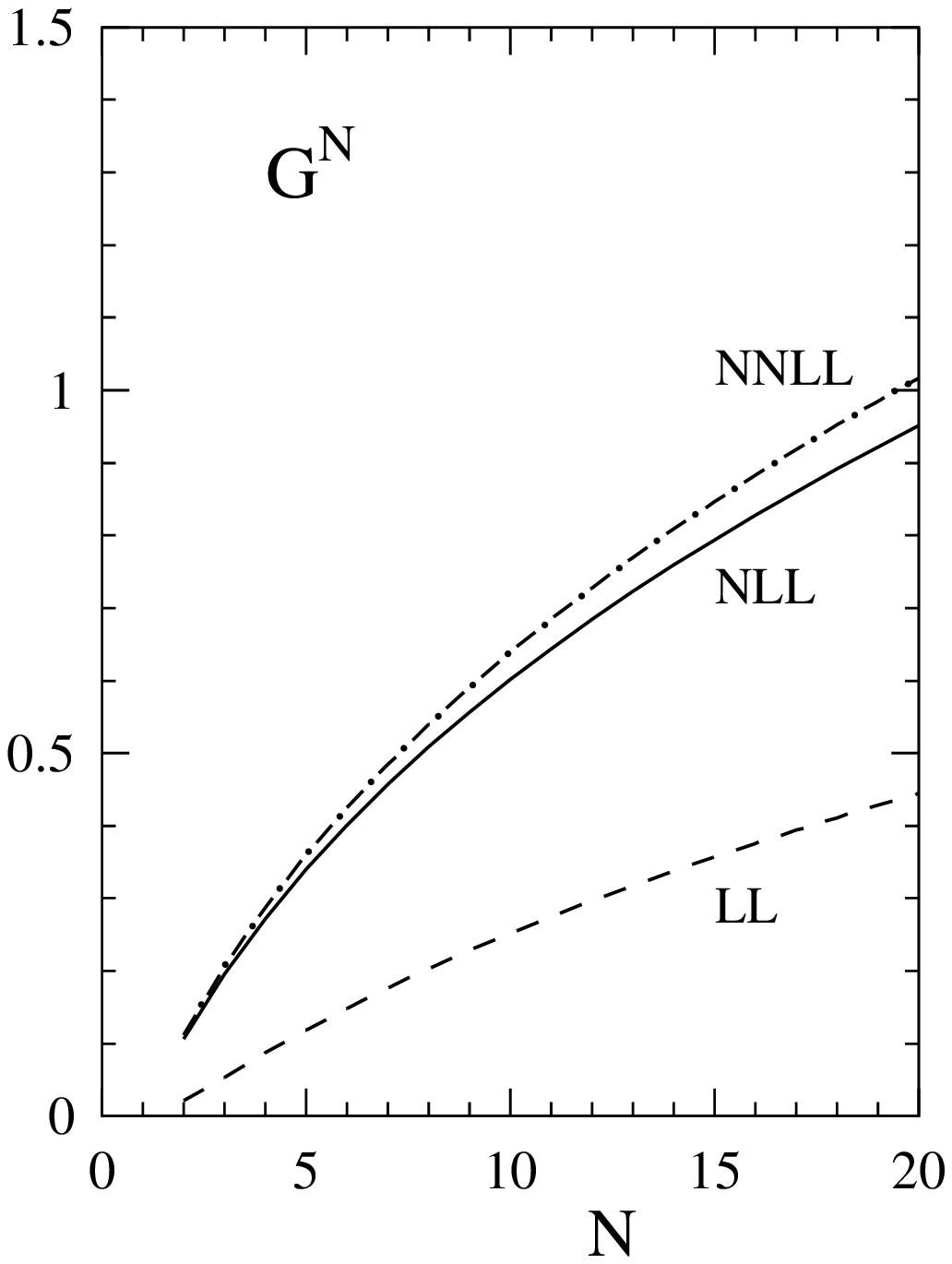}
\epsfxsize=6.2cm\epsfbox{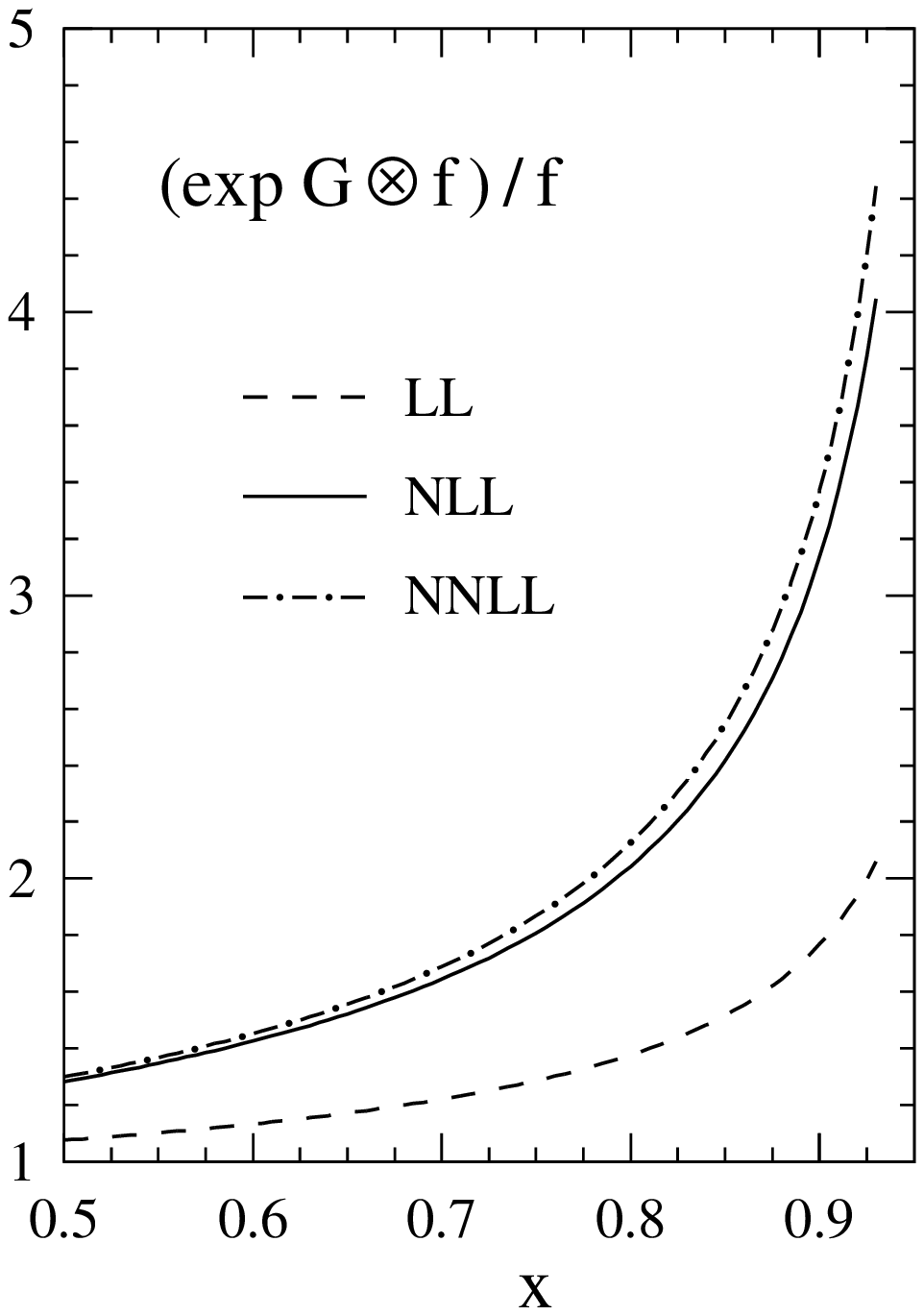}
}
\vspace*{-0.2cm}
\caption{\label{fig:one}
Left: 
 The LL, NLL and NNLL approximations for the resummation
 exponent $G^N(Q^2)$ in eq.~(\ref{eq:GNexp}) at $\mu_r^2 = \mu_f^2 = Q^2$ 
 for $\as(Q^2) = 0.2$ and four flavours. 
Right: These results convoluted with a typical input shape 
 $xf = x^{1/2}(1-x)^3$.}
\end{center}
\vspace*{-0.5cm}
\end{figure}


\begin{thebibliography}{10}

\bibitem{vanNeerven:1991nn}
W.~L. van Neerven and E.~B. Zijlstra,
\newblock Phys. Lett. {\bf B272}, 127 (1991)

\bibitem{Zijlstra:1991qc}
E.~B. Zijlstra and W.~L. van Neerven,
\newblock Phys. Lett. {\bf B273}, 476 (1991)

\bibitem{Zijlstra:1992kj}
E.~B. Zijlstra and W.~L. van Neerven,
\newblock Phys. Lett. {\bf B297}, 377 (1992)

\bibitem{Zijlstra:1992qd}
E.~B. Zijlstra and W.~L. van Neerven,
\newblock Nucl. Phys. {\bf B383}, 525 (1992)

\bibitem{Moch:1999eb}
S.~Moch and J.~A.~M. Vermaseren,
\newblock Nucl. Phys. {\bf B573}, 853 (2000)

\bibitem{Larin:1994vu}
S.~A. Larin, T.~van Ritbergen, and J.~A.~M. Vermaseren,
\newblock Nucl. Phys. {\bf B427}, 41 (1994)

\bibitem{Larin:1997wd}
S.~A. Larin, P.~Nogueira, T.~van Ritbergen, and J.~A.~M. Vermaseren,
\newblock Nucl. Phys. {\bf B492}, 338 (1997)

\bibitem{Retey:2000nq}
A.~Retey and J.~A.~M. Vermaseren,
\newblock Nucl. Phys. {\bf B604}, 281 (2001)

\bibitem{Moch:2002sn}
S.~Moch, J.~A.~M. Vermaseren, and A.~Vogt,
\newblock Nucl. Phys. {\bf B646}, 181 (2002)

\bibitem{Sterman:1987aj}
G.~Sterman,
\newblock Nucl. Phys. {\bf B281}, 310 (1987)

\bibitem{Catani:1989ne}
S.~Catani and L.~Trentadue,
\newblock Nucl. Phys. {\bf B327}, 323 (1989)

\bibitem{Catani:1991rp}
S.~Catani and L.~Trentadue,
\newblock Nucl. Phys. {\bf B353}, 183 (1991)

\bibitem{Vogt:2000ci}
A.~Vogt,
\newblock Phys. Lett. {\bf B497}, 228 (2001)

\bibitem{Contopanagos:1997nh}
H.~Contopanagos, E.~Laenen, and G.~Sterman,
\newblock Nucl. Phys. {\bf B484}, 303 (1997)

\bibitem{vanNeerven:2000wp}
W.~L. van Neerven and A.~Vogt,
\newblock Phys. Lett. {\bf B490}, 111 (2000)

\bibitem{Blumlein:1995jp}
J.~Bl\"umlein and A.~Vogt,
Phys.\ Lett.\ B {\bf 370}, 149 (1996)

\bibitem{Gracey:nn}
J.~A.~Gracey,
Phys.\ Lett.\ B {\bf 322}, 141 (1994)

\bibitem{Berger:2002sv}
C.~F. Berger,
\newblock Phys. Rev. {\bf D66}, 116002 (2002)

\bibitem{Matsuura:1989sm}
T.~Matsuura, S.~C. van~der Marck, and W.~L. van Neerven,
\newblock Nucl. Phys. {\bf B319}, 570 (1989)

\bibitem{Forte:2002ni}
S.~Forte and G.~Ridolfi,
\newblock Nucl. Phys. {\bf B650}, 229 (2003)

\bibitem{Gardi:2002xm}
E.~Gardi and R.~G. Roberts,
\newblock Nucl. Phys. {\bf B653}, 227 (2003)

\bibitem{Catani:1996yz}
S.~Catani, M.~L. Mangano, P.~Nason, and L.~Trentadue,
\newblock Nucl. Phys. {\bf B478}, 273 (1996)

\bibitem{vanNeerven:2001pe}
W.~L.~van Neerven and A.~Vogt,
Nucl.\ Phys.\ B {\bf 603}, 42 (2001)

\end{thebibliography}
\end{document}